\documentclass[onecolomn, journal]{IEEEtran}
\usepackage{multirow}
\usepackage{amsmath}
\usepackage{graphicx}
\usepackage{epstopdf}
\usepackage{bm}
\usepackage{amsfonts}
\usepackage{subfigure}
\usepackage[colorlinks,linkcolor=black,anchorcolor=black,citecolor=black]{hyperref}
\usepackage{cite}
\usepackage{makecell}
\usepackage{booktabs}

\makeatletter
\newif\if@restonecol
\makeatother

\usepackage[ruled,vlined]{algorithm2e}
\usepackage{algpseudocode}
\usepackage{amsmath}
\mathchardef\mhyphen="2D

\ifCLASSINFOpdf
\else
\fi

\begin{document}
%
\title{Multi-task Deep Neural Networks for Massive MIMO CSI Feedback}
%
%
%

\author{Boyuan~Zhang,
        Haozhen~Li,
        Xin~Liang,
        Xinyu~Gu,~\IEEEmembership{Member,~IEEE}, 
        and~Lin~Zhang,~\IEEEmembership{Member,~IEEE}
\thanks{This work was supported in part by the State Major Science and Technology Special Projects (Grant No.2018ZX03001024).}
\thanks{B. Zhang, H. Li, X. Liang, L. Zhang are with the School of Artificial Intelligence, Beijing University of Posts and Telecommunications, Beijing,
100876, China (e-mail: \{zhangboyuan, lihaozhen, liangxin, zhanglin\}@bupt.edu.cn).

X. Gu is with the School of Artificial Intelligence, Beijing University of Posts and Telecommunications, Beijing,
100876, China. She is also with the Purple Mountain Laboratories, Nanjing 211111, China (e-mail: guxinyu@bupt.edu.cn).

This work has been submitted to the IEEE for possible publication. Copyright may be transferred without notice, after which this version may no longer be accessible.
}
}

\maketitle

\begin{abstract}
Deep learning has been widely applied for the channel state information (CSI) feedback in frequency division duplexing (FDD) massive multiple-input multiple-output (MIMO) system.
For the typical supervised training of the feedback model, the requirements of large amounts of task-specific labeled data can hardly be satisfied, and the huge training costs and storage usage of the model in multiple scenarios are hindrance for model application.
In this letter, a multi-task learning-based approach is proposed to improve the feasibility of the feedback network. An encoder-shared feedback architecture and the corresponding training scheme are further proposed to facilitate the implementation of the multi-task learning approach.
The experimental results indicate that the proposed multi-task learning approach can achieve comprehensive feedback performance with considerable reduction of training cost and storage usage of the feedback model.
\end{abstract}


%
\IEEEpeerreviewmaketitle

\section{Introduction}

\IEEEPARstart{M}{assive} multiple-input multiple-output (MIMO) has been regarded as a crucial technology in the fifth generation (5G) wireless communications \cite{MIMO}. The acquisition of downlink channel state information (CSI) at the base station (BS) is required to achieve the performance gain. Due to the lack of channel reciprocity in the frequency division duplexing (FDD) mode, the downlink CSI has to be fed back from the user equipment (UE) to the BS. The excessive feedback overhead caused by the large-scale antennas makes it hard to achieve accurate and efficient CSI feedback, which has become a major barrier for the application of MIMO technology.

The compressive sensing (CS) \cite{CS} feedback method relies on a simple sparsity prior which is not satisfied by the real channels and the time cost is also unaffordable. With the wide applications of deep learning (DL) in wireless communications, the DL-based approach has been put forward to cope with the challenges in CSI feedback. CsiNet \cite{CsiNet} proposes the autoencoder-based structure to accomplish the CSI compression and reconstruction, which provides a basic approach for the DL-based CSI feedback research. Some works including DualNet \cite{DualNet}, ATNet \cite{ATNet} and CRNet \cite{CRNet} focus on the improvement of feedback accuracy through the novel designs of network structures, while some other researches such as CVLNet \cite{CVLNet} and BCsiNet \cite{BCsiNet} take the complexity and feasibility of the model into consideration.

Most existing researches implement the supervised singe-task learning approach. Task-specific network models are trained for different channel scenarios. To achieve acceptable performance, large-scale dataset is required for model training in each scenario, causing low efficiency in time and memory. Different network models have to be stored both at the UE and the BS to ensure the feedback accuracy in different scenarios, which leads to the storage problem, especially at the UE. The transfer learning and meta learning methods \cite{DTL-MAML} utilized the cross-task relativity in multiple scenarios to reduce the training cost and the size of the required dataset, while the problem of model complexity still exists.

In this letter, the multi-task learning approach is applied to the DL-based CSI feedback in the massive MIMO system. By utilizing the labeled data from related tasks, the amount of training data can be reduced, contributing to the savings of training costs. The more general representations of the CSI matrices in different scenarios can be learned and a shared network architecture can be further constructed to lower the total parameters of the model, which is helpful to solve the storage problem.

The multi-task learning strategy is first introduced to clarify the training of the CSI feedback model. Only small-size dataset is required in each scenario for model training, and some layers of model can be reused in multiple scenarios, which contributes to the storage saving at the UE. Then an encoder-shared architecture is further designed for the autoencoder-based CSI feedback network and the corresponding training scheme is also provided. According to the experimental results based on the COST 2100 \cite{COST-2100} and Clustered Delay Line (CDL) \cite{38.901} channel models, the size of training set and time cost can be considerably reduced and comprehensive feedback accuracy can be achieved.

The rest of the letter is organized as follows. The system model is introduced in Section \uppercase\expandafter{\romannumeral2}. The multi-task learning strategy, encoder-shared architecture and training scheme are demonstrated in Section \uppercase\expandafter{\romannumeral3}. Experimental settings and results are presented in Section \uppercase\expandafter{\romannumeral4} and the letter is concluded in Section \uppercase\expandafter{\romannumeral5}.

\begin{figure*}[t]
    \centering
        \includegraphics[width=6.4 in]{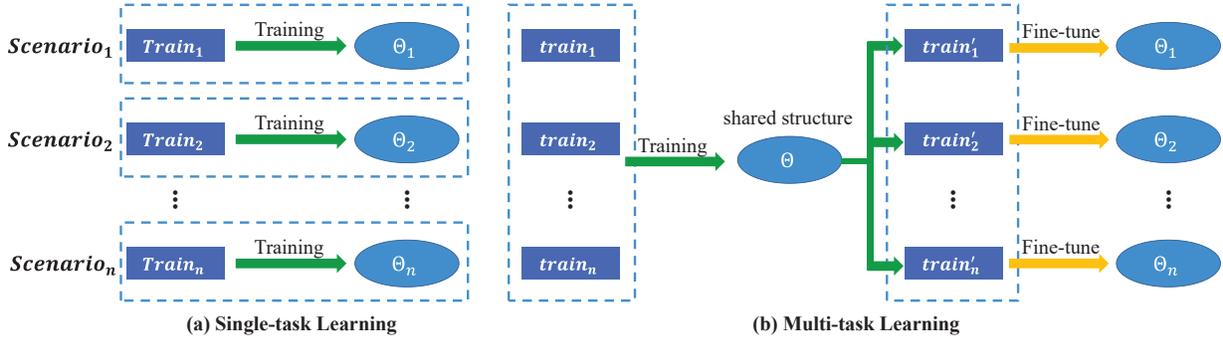}
        \caption{Comparison between the single-task and multi-task learning strategies.}
    \label{Strategy}
\end{figure*}

\section{System Model}

In this letter, we consider a single-cell downlink massive MIMO system under OFDM mode with $\tilde{N}_{c}$ subcarriers studied in \cite{CsiNet}. There are $N_{t}\gg1$ transmitting antennas configured at the BS and a single antenna configured at the UE. The received signal at the \emph{n}th subcarrier can be presented as:

\begin{equation}
y_n = \tilde{\textbf{h}}_{n}^{H}\textbf{v}_{n}x_{n} + z_{n}.
\end{equation}

For the \emph{n}th subcarrier, $\tilde{\textbf{h}}_{n} \in \mathbb{C}^{N_{t} \times 1}$, $\textbf{v}_{n} \in \mathbb{C}^{N_{t} \times 1}$, $x_n \in \mathbb{C}$, $z_{n} \in \mathbb{C}$ stand for channel frequency response, precoding vector, modulated data symbol and additive noise, respectively. The downlink CSI matrix in the spatial frequency domain can be denoted as $\tilde{\textbf{H}}=[\tilde{\textbf{h}}_{1},\tilde{\textbf{h}}_{2},...,\tilde{\textbf{h}}_{\tilde{N}_{c}}]^{H} \in \mathbb{C}^{\tilde{N}_{c} \times N_{t}}$. The BS can design the precoding vectors and further boost the communication quality of the massive MIMO system according to the received $\tilde{\textbf{H}}$.

However, the CSI matrix consists of $N_{t}\tilde{N}_{c}$ complex numbers, which leads to the huge feedback overhead. By using the 2D discrete Fourier transform (DFT), the CSI matrix $\tilde{\textbf{H}}$ can be converted into the angular-delay domain as follows:

\begin{equation}
\textbf{H} = \textbf{F}_{\rm d}\tilde{\textbf{H}}\textbf{F}_{\rm a}^{H},
\end{equation}

\noindent where $\textbf{F}_{\rm d} \in \mathbb{C}^{\tilde{N}_{c} \times \tilde{N}_{c}}$ and $\textbf{F}_{\rm a} \in \mathbb{C}^{N_{t} \times N_{t}}$  are DFT matrices. Most of the values in the rows of $\textbf{H}$ are close to zero because of the channel sparsity in the angular-delay domain, the channel matrix is truncated to $N_{t} \times N_{c}$ and further used for the feedback process.

The deep learning-based scheme provide a new approach for CSI feedback by training the autoencoder-based neural network to achieve the compression and reconstruction of the CSI matrix. The encoder at the UE first compresses the CSI matrix H into a codeword with a lower dimension as follows:

\begin{equation}
\textbf{s} = f_{\rm en}(\textbf{H}).
\end{equation}

Then the codeword is transmitted through the feedback channel to the BS. The CSI reconstruction is then performed by the decoder at the BS, which is presented as:

\begin{equation}
\hat{\textbf{H}} = f_{\rm de}(\textbf{s}),
\end{equation}

\noindent where $\hat{\textbf{H}}$ is the reconstrued CSI matrix. Through training the autoencoder-based network to minimize the difference between $\textbf{H}$ and $\hat{\textbf{H}}$, effective and accurate CSI feedback can be achieved.

\section{Multi-task Learning for CSI Feedback}

In this section, the multi-task learning approach for CSI feedback is demonstrated in detail.
The multi-task learning strategy is first introduced and the differences between the single-task and multi-task learning approaches are also provided. Then the encoder-shared architecture is proposed as the application structure of multi-task learning on CSI feedback network.
Finally, the training scheme is designed for the multi-task learning approach.

\subsection{Multi-task learning strategy}

In most of the existing DL-based CSI feedback researches, multiple scenarios are considered in the training and testing of the network models, such as $Indoor$ and $Outdoor$ in COST 2100 and $CDL-A/B/C/D/E$ in CDL channel model. Not only the real-time changes of the channel environment, but also the movement of the mobile devices will lead to the changes of the channel scenario. Therefore, the learning of the network in one scenario can be regarded as a specific task. The relatedness of different tasks will be utilized under the multi-task approach. Fig. \ref{Strategy} shows the difference between the single-task and multi-task strategies in $n$ scenarios.

The single-task learning strategy is shown in Fig. \ref{Strategy}(a). In $Scenario_i$, the network model $\Theta_{i}$ is trained using the large-size training set $Train_i$. The training of the models in different scenarios are independent. For the application of CSI feedback in multiple scenarios, the large-size dataset in every scenario including $Train_1$, $Train_2$, ... $Train_n$ should be obtained and thousands of iterations are required in the training stage. Different network models including $\Theta_{1}$, $\Theta_{2}$, ... $\Theta_{n}$ should be stored at the UE and the BS in the inference stage. Therefore, the high training and memory costs become a hindrance for the application of the feedback network in the real system.

\begin{figure*}[t]
    \centering
        \includegraphics[width=5 in]{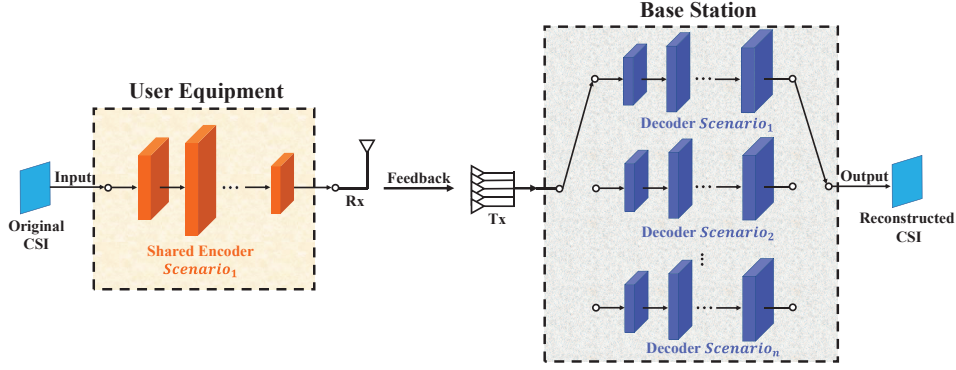}
        \caption{Encoder-shared architecture.}
    \label{encoder-shared}
\end{figure*}

The multi-task learning strategy is shown in Fig. \ref{Strategy}(b). Based on a shared structure, the training of $n$ tasks in $n$ scenarios are first performed to capture the intrinsic relatedness between different tasks, which is often achieved by training a general model $\Theta$ with the combined dataset composed of the small-size training set $train_i$ in each task. This is an effective way to leverage the supervised data from related tasks. Then, the specific layers of the shared structure, or the additional layers apart from the shared structure will be fine-tuned with specific small-size training data $train^{'}_{i}$ in a specific task $Scenario_i$ to ensure the convergence of the model in each task.

By applying the multi-task strategy, the size of the training set required in each scenario is considerably decreased and the training cost can be further reduced. The shared layers can be reused in multiple scenarios, which is helpful to alleviate the storage problem, especially at the UE.

\subsection{Encoder-shared architecture}

The autoencoder-based deep neural network is used in the DL-based CSI feedback researches. The bottleneck of the application of the network in practical systems mainly occurs at the UE side. The model complexity, which is often measured by the total parameters should be considered due to limitation of storage space of the mobile device. For single-task learning, $n$ sets of parameters of the encoder need to be stored at the UE for the feature extraction and data compression of the input CSI in $n$ scenarios, which hinders the application of the feedback network in real massive MIMO systems.

\begin{algorithm}[!b]
\caption{Training scheme of the network models using encoder-shared architecture.}

\textbf{\emph{Pre-training:}}

\KwIn{$train_{multi}$ composed of the small-size datasets $train_{1}$, $train_{2}$, ... $train_{n}$;
      Number of gradient descents for pre-training: $epoch_1$.}
\KwOut{Parameters of the pre-trained general model $\Theta$, including encoder $\theta_{en}$ and decoder $\theta_{de}$.}

\For{ $i = 1, 2, ... epoch_1$ }
{
    Update the parameters using $train_{multi}$;
}
Save the parameters $\theta_{en}$ and $\theta_{de}$;

\textbf{\emph{Fine-tuning:}}

\KwIn{Smaller sized datasets $train^{'}_{1}$, $train^{'}_{2}$, ... $train^{'}_{n}$;
      Parameters of the pre-trained general model $\Theta$, including encoder $\theta_{en}$ and decoder $\theta_{de}$;
      Number of gradient descents for fine-tuning: $epoch_2$.}
\KwOut{Parameters of the fine-tuned task-specific decoder $\theta^{'}_{de\_i}$.}

\For{ $Scenario = 1, 2, ... n$ }
{
    Load $\theta_{en}$ and $\theta_{de}$;\\
    Set $\theta_{en}$ as untrainable;\\
    \For{ $i = 1, 2, ... epoch_2$ }
    {
        Update the parameters using $train^{'}_{i}$;
    }
    Save the parameters $\theta^{'}_{de\_i}$;
}

\textbf{\emph{Test:}}

\KwIn{Test sets $test_{1}$, $test_{2}$, ... $test_{n}$;
      Parameters of the pre-trained encoder $\theta_{en}$ and the fine-tuned decoder $\theta^{'}_{de\_1}$, $\theta^{'}_{de\_2}$, ... $\theta^{'}_{de\_n}$;}
\KwOut{Test results measured by NMSE in dB.}

\For{ $Scenario = 1, 2, ... n$ }
{
    Load $\theta_{en}$ and $\theta^{'}_{de\_i}$;\\
    Predict the reconstructed CSI and calculate the NMSE using $test_i$;
}

\end{algorithm}

The parameter sharing of the hidden layers is the most commonly used way to perform multi-task learning in deep neural networks. Therefore, based on the multi-task strategy, an encoder-shared architecture is proposed to solve the above storage problem at the UE, as shown in Fig. \ref{encoder-shared}. For the application of the encoder-shared architecture in $n$ tasks, i.e. $n$ scenarios, the encoder is shared in different tasks, while the decoder is task-specific. Therefore, only one set of parameters of the encoder is required for the $n$ scenarios, which considerably alleviate the burden of memory at the UE.

The encoder-shared architecture is applicable to most of the autoencoder-based feedback networks proposed in the existing studies. For the application of the network with encoder-shared architecture in $n$ scenarios, the parameters at the UE will be reduced to $1/n$ compared with the typical task-specific architecture.

\subsection{Training scheme}

Based on the multi-task learning strategy, the training scheme of the network models using encoder-shared architecture is proposed, as shown in \textbf{Algorithm 1}. The pre-training and fine-tuning procedures will be performed to obtain the feedback models in $n$ scenarios.

During the pre-training procedure, the combined dataset $train_{multi}$ is first obtained, which is composed of the small-size datasets $train_i$ in the $n$ scenarios. Then a single general model $\Theta$ including the encoder $\theta_{en}$ and the decoder $\theta_{de}$ will be trained with $train_{multi}$.

After obtaining the general model $\Theta$, the parameters of the encoder $\theta_{en}$ will be used in each scenario. Then for the feedback task in $Scenario_i$, the autoencoder will be fine-tuned using a smaller sized dataset $train^{'}_{i}$ in $Scenario_i$, with the encoder set as untrainable. The specific decoder model $\theta^{'}_{de\_i}$ can be obtained, which is still coordinated with the encoder $\theta_{en}$. The feedback of the CSI in $Scenario_i$ will be achieved with the general encoder $\theta_{en}$ and the fine-tuned decoder $\theta^{'}_{de\_i}$.

The general model $\Theta$ obtained by the pre-training using combined dataset learns the general representations of the CSI data, which is able converge faster in the specific task. So the epoch for fine-tuning $epoch_2$ is smaller than the epoch for pre-training $epoch_1$, which is helpful to shorten the total training time.

\section{Simulation Results and Analysis}

\subsection{Data settings}

\begin{table}[b]
\renewcommand{\arraystretch}{1.2}
\caption{Basic parameter settings of data generation}
\label{Settings}
\centering
\begin{tabular}{c|c}
\hline
Parameters                       &   Values                \\
\hline
Delay profile                    &   CDL-A, B, C, D, E \\
\hline
Downlink Carrier frequency       &   2.1 GHz\\
\hline
Subcarrier spacing               &  15 kHz\\
\hline
Number of subcarriers            &   72\\
\hline
OFDM symbols                     &   14\\
\hline
Transmitting antennas            &   32 (Uniform Linear Array)\\
\hline
Receiving antennas               &   2\\
\hline
Velocities of UE                 &   4.8, 24, 40, 60 km/h\\
\hline
\end{tabular}
\end{table}

The performance of the proposed multi-task learning approach is evaluated based on the COST 2100\cite{COST-2100} and the nrCDLChannel model formulated in 3GPP TR 38.901 \cite{38.901}.

For COST 2100, the same dataset in \cite{CsiNet} is used in this letter. There are 100,000 training, 30,000 validation and 20,000 testing samples in the Indoor and Outdoor scenarios respectively.

For CDL, the datasets in the five channel scenarios including CDL-A, CDL-B, CDL-C, CDL-D, CDL-E are generated. Some basic parameter settings of data generation are shown in Table \ref{Settings}. For CDL-A, 60,000 samples are generated, including 50,000 training, 5,000 validation and 5,000 testing samples. For the CDL-B, C, D, E, 10,000 samples are generated in each scenario, including 4,000 training, 1,000 validation and 5,000 testing samples.


\subsection{Experiment settings}

To evaluate the multi-task learning approach, two sets of experiments are performed based on the COST 2100 and CDL channel models respectively.

COST 2100: The experiments of multi-task learning approach are implemented on the CsiNet \cite{CsiNet} and CRNet \cite{CRNet}. In the pre-training procedure, $train_{multi}$ is composed of 50,000 training, 15,000 validation samples from the Indoor scenario ($train_{in}$) and 50,000 training, 15,000 validation from Outdoor scenario ($train_{out}$). The learning rate is 0.001, batch size is 200, and $epoch_1$ is 1,000. In the fine-tuning procedure, $train^{'}_{in}$ and $train^{'}_{out}$ are composed of 25,000 training, 15,000 validation samples respectively, which are from $train_{in}$ and $train_{out}$.  The learning rate is 0.0001, batch size is 200, and $epoch_2$ is 500. The experimental results presented in CsiNet \cite{CsiNet} and CRNet \cite{CRNet} are used for comparison.

CDL: According to the research \cite{DTL-MAML} based on CDL channel model, transfer learning approaches is able to achieve comprehensive performance with lower training cost. Therefore, the experiment of transfer learning is first performed as the comparison.
The same network structure and experimental setting in \cite{DTL-MAML} is applied. The experiment of multi-task learning is also performed on the network proposed in \cite{DTL-MAML}. In the pre-training procedure, $train_{multi}$ is composed of 4,000 training, 1,000 validation samples of CDL-A, B, C, D, E ($train_{i}$) respectively. The learning rate is 0.001, batch size is 50, and $epoch_1$ is 200. In the fine-tuning procedure, $train^{'}_{i}$ is composed of 2,000 training, 500 validation samples, which are from $train_{i}$. The learning rate is 0.0001, batch size is 50, and $epoch_2$ is 100.

For the network training in the above experiments, Mean Square Error (MSE) is used as the loss function and the Adam optimizer is adopted to update the parameters. Normalized Mean Square Error (NMSE) is used to evaluate the accuracy of CSI reconstruction as (6). The training and testing processes are accomplished using Nvidia GeForce RTX 2080 Ti GPU.

\begin{equation}
NMSE = \mathbb{E}\{\frac{||\textbf{H}-\hat{\textbf{H}}||_{2}^{2}}{||\textbf{H}||_{2}^{2}}\}.
\end{equation}

\subsection{Results and analysis}

\begin{table}[b]
\renewcommand{\arraystretch}{1.2}
\caption{Comparison of the feedback performance based on COST 2100}
\label{Result_COST}
\centering
\begin{tabular}{c|c|c|c|c}
\hline
\multirow{3}{0.5cm}{CR}   &                              \multicolumn{4}{c}{CsiNet}                             \\
\cline{2-5}
                          &          \multicolumn{2}{c|}{Indoor}     &          \multicolumn{2}{c}{Outdoor}      \\
\cline{2-5}
                          &     Single-task    &      Multi-task     &      Single-task    &      Multi-task     \\
\hline
1/4                        &   -17.36          &    -19.28           &        -8.75        &         -11.57      \\
1/16                       &   -8.65           &    -8.84            &        -4.51        &         -4.83       \\
1/32                       &   -6.24           &    -6.33            &        -2.81        &         -2.88       \\
1/64                       &   -5.84           &    -5.00            &        -1.93        &         -2.05       \\
\hline
\hline
\multirow{3}{0.5cm}{CR}   &                              \multicolumn{4}{c}{CRNet}                             \\
\cline{2-5}
                          &        \multicolumn{2}{c|}{Indoor}        &       \multicolumn{2}{c}{Outdoor}         \\
\cline{2-5}
                          &     Single-task    &      Multi-task      &      Single-task   &      Multi-task     \\
\hline
1/4                        &   -21.17           &    -22.67           &        -10.42      &        -11.70      \\
1/16                       &   -10.29           &    -10.68           &        -5.09       &         -5.47      \\
1/32                       &   -8.58            &    -8.13            &        -3.19       &         -3.10      \\
1/64                       &   -6.14            &    -5.82            &        -2.13       &         -2.29      \\
\hline
\end{tabular}
\end{table}

\begin{table}[t]
\renewcommand{\arraystretch}{1.2}
\caption{Comparison of the feedback performance based on CDL}
\label{Result_CDL}
\centering
\begin{tabular}{m{1cm}<{\centering}|m{1.2cm}<{\centering}|m{1.4cm}<{\centering}|m{1.4cm}<{\centering}}
\hline
\multirow{2}{0.5cm}{CR}  &   \multirow{2}{0.9cm}{Scenario}    & \multicolumn{2}{c}{Fully Convolutional}    \\
\cline{3-4}
                         &                                    &   Transfer    &   Multi-task        \\
\hline
\multirow{5}{10pt}{1/8}
                         &              CDL-A                 &  -22.84        &    -22.84    \\
                         &              CDL-B                 &  -5.85         &    -14.99    \\
                         &              CDL-C                 &  -11.98        &    -20.90    \\
                         &              CDL-D                 &  -11.93        &    -21.27    \\
                         &              CDL-E                 &  -13.89        &    -21.03    \\
\hline
\hline
\multirow{5}{10pt}{1/32}
                         &              CDL-A                 &  -22.84        &    -22.84    \\
                         &              CDL-B                 &  -5.85         &    -14.99    \\
                         &              CDL-C                 &  -11.98        &    -20.90    \\
                         &              CDL-D                 &  -11.93        &    -21.27    \\
                         &              CDL-E                 &  -13.89        &    -21.03    \\
\hline
\hline
\multirow{5}{10pt}{1/64}
                         &              CDL-A                 &  -22.84        &    -22.84    \\
                         &              CDL-B                 &  -5.85         &    -14.99    \\
                         &              CDL-C                 &  -11.98        &    -20.90    \\
                         &              CDL-D                 &  -11.93        &    -21.27    \\
                         &              CDL-E                 &  -13.89        &    -21.03    \\
\hline
\end{tabular}
\end{table}

Experiment results of the feedback performance measured by NMSE in dB are analyzed in this part. The results based on COST 2100 at the compression rates of 1/4, 1/16, 1/32 and 1/64 in the Indoor and Outdoor scenarios are shown in Table \ref{Result_COST}. The results based on CDL at the compression rates of 1/8, 1/32 and 1/64 in CDL-A, B, C, D, E scenarios are presented in Table \ref{Result_CDL}.

Compared with the results of single-task learning experiment on COST 2100 channel model and transfer learning experiment on CDL channel model, the network model based on the multi-task approach with shared-encoder architecture can achieve comparable feedback performance in different scenarios. In some cases, the performance of multi-task learning approach is better than the single-task and transfer learning approaches.

To better compare the feasibility of different approaches, the model complexity measured by parameters at the UE and the training cost measured by the number of training samples and total training time are listed in Table \ref{Complexity}, which takes the results of $CR=1/4$ for COST 2100 and $CR=1/8$ for CDL as examples.

\begin{table}[b]
\renewcommand{\arraystretch}{1.2}
\caption{Comparison of the model complexity and training cost based on COST 2100 and CDL}
\label{Complexity}
\centering
\begin{tabular}{c|c|c|c}
\hline
Training         &         \multicolumn{3}{c}{CsiNet}         \\
\cline{2-4}
Methods          &    Parameters (UE)     &    Size of training set     &   Training time      \\
\hline
Single-task      &     2,098,508          &        200,000              &     13.88h           \\
Multi-task       &     1,049,254          &        100,000              &     6.67h            \\
\hline
Reduction        &        50\%            &          50\%               &     51.95\%          \\
\hline
\hline
Training         &         \multicolumn{3}{c}{CRNet}         \\
\cline{2-4}
Methods          &    Parameters (UE)     &    Size of training set     &   Training time      \\
\hline
Single-task      &     2,099,780          &        200,000              &     13.88h           \\
Multi-task       &     1,049,890          &        100,000              &     6.94h            \\
\hline
Reduction        &        50\%            &          50\%               &     50\%             \\
\hline
\hline
Training         &         \multicolumn{3}{c}{CDL}         \\
\cline{2-4}
Methods          &    Parameters (UE)     &    Size of training set     &   Training time      \\
\hline
Transfer         &     2,099,780          &        200,000              &     13.88h           \\
Multi-task       &     1,049,890          &        100,000              &     6.94h            \\
\hline
Reduction        &        50\%            &          50\%               &     50\%             \\
\hline
\end{tabular}
\end{table}

For training cost, single-task learning requires large-size dataset in the Indoor and Outdoor scenarios (100,000 training), and transfer learning requires large-size training dataset in the pre-trained CDL-A scenario (50,000 training), and small-size datasets in CDL-B, C, D, E scenarios (5,000 training each). For multi-task learning, only small-size dataset (50,000 training for Indoor and Outdoor in COST 2100, 5,000 training for CDL-A, B, C, D, E in CDL) is required for each scenario. 50\% reduction of training samples can be achieved, which is more feasible in the real system and further contributes to the saving of total training time (about 50\%).

For model complexity, 2 network models for COST 2100 and 5 network models for CDL need to be stored at the UE and the BS using single-task and transfer approaches. While for multi-task learning, only one general encoder model is required, leading to 50\% (COST 2100) and 80\%(CDL) reduction of parameters at the UE, which is of great significance to alleviate the storage problem.

Therefore, the multi-task learning approach is able to achieve considerable reduction of model complexity and training cost with the guarantee of feedback accuracy, which improves the feasibility of the feedback model in real massive MIMO systems

\section{Conclusion}

In this letter, the multi-task learning approach is introduced and applied on CSI feedback in the massive MIMO system. The multi-task learning strategy, encoder-shared network architecture and the corresponding training scheme are demonstrated in detail. The experimental results based on the COST 2100 and CDL channel model indicate that the multi-task learning approach can achieve comprehensive feedback performance with considerable reduction of model complexity and training cost, which is helpful to improve the feasibility of the feedback network in the practical massive MIMO systems.

\bibliographystyle{IEEEtran}
\bibliography{IEEEexample}

\begin{thebibliography}{10}
\providecommand{\url}[1]{#1}
\csname url@samestyle\endcsname
\providecommand{\newblock}{\relax}
\providecommand{\bibinfo}[2]{#2}
\providecommand{\BIBentrySTDinterwordspacing}{\spaceskip=0pt\relax}
\providecommand{\BIBentryALTinterwordstretchfactor}{4}
\providecommand{\BIBentryALTinterwordspacing}{\spaceskip=\fontdimen2\font plus
\BIBentryALTinterwordstretchfactor\fontdimen3\font minus
  \fontdimen4\font\relax}
\providecommand{\BIBforeignlanguage}[2]{{%
\expandafter\ifx\csname l@#1\endcsname\relax
\typeout{** WARNING: IEEEtran.bst: No hyphenation pattern has been}%
\typeout{** loaded for the language `#1'. Using the pattern for}%
\typeout{** the default language instead.}%
\else
\language=\csname l@#1\endcsname
\fi
#2}}
\providecommand{\BIBdecl}{\relax}
\BIBdecl

\bibitem{MIMO}
M.~Agiwal, A.~Roy, and N.~Saxena, ``Next generation {5G} wireless networks: A
  comprehensive survey,'' \emph{IEEE Communications Surveys Tutorials},
  vol.~18, no.~3, pp. 1617--1655, 2016.

\bibitem{CS}
P.-H. Kuo, H.~T. Kung, and P.-A. Ting, ``Compressive sensing based channel
  feedback protocols for spatially-correlated massive antenna arrays,'' in
  \emph{2012 IEEE Wireless Communications and Networking Conference (WCNC)},
  2012, pp. 492--497.

\bibitem{CsiNet}
C.-K. Wen, W.-T. Shih, and S.~Jin, ``Deep learning for massive {MIMO} {CSI}
  feedback,'' \emph{IEEE Wireless Communications Letters}, vol.~7, no.~5, pp.
  748--751, 2018.

\bibitem{DualNet}
Z.~Liu, L.~Zhang, and Z.~Ding, ``Exploiting bi-directional channel reciprocity
  in deep learning for low rate massive {MIMO} {CSI} feedback,'' \emph{IEEE
  Wireless Communications Letters}, vol.~8, no.~3, pp. 889--892, 2019.

\bibitem{ATNet}
H.~Chang, X.~Liang, H.~Li, J.~Shen, X.~Gu, and L.~Zhang, ``Deep learning-based
  bitstream error correction for {CSI} feedback,'' \emph{IEEE Wireless
  Communications Letters}, vol.~10, no.~12, pp. 2828--2832, 2021.

\bibitem{CRNet}
Z.~Lu, J.~Wang, and J.~Song, ``Multi-resolution {CSI} feedback with deep
  learning in massive {MIMO} system,'' in \emph{ICC 2020 - 2020 IEEE
  International Conference on Communications (ICC)}, 2020, pp. 1--6.

\bibitem{CVLNet}
H.~Li, B.~Zhang, H.~Chang, X.~Liang, and X.~Gu, ``{CVLN}et: A complex-valued
  lightweight network for {CSI} feedback,'' \emph{IEEE Wireless Communications
  Letters}, pp. 1--1, 2022.

\bibitem{BCsiNet}
Z.~Lu, J.~Wang, and J.~Song, ``Binary neural network aided {CSI} feedback in
  massive {MIMO} system,'' \emph{IEEE Wireless Communications Letters},
  vol.~10, no.~6, pp. 1305--1308, 2021.

\bibitem{DTL-MAML}
Y.~Yang, F.~Gao, Z.~Zhong, B.~Ai, and A.~Alkhateeb, ``Deep transfer
  learning-based downlink channel prediction for {FDD} massive {MIMO}
  systems,'' \emph{IEEE Transactions on Communications}, vol.~68, no.~12, pp.
  7485--7497, 2020.

\bibitem{COST-2100}
L.~Liu, C.~Oestges, J.~Poutanen, K.~Haneda, P.~Vainikainen, F.~Quitin,
  F.~Tufvesson, and P.~D. Doncker, ``The {COST} 2100 {MIMO} channel model,''
  \emph{IEEE Wireless Communications}, vol.~19, no.~6, pp. 92--99, 2012.

\bibitem{38.901}
3GPP, ``{Study on channel model for frequencies from 0.5 to 100 GHz},'' {3rd
  Generation Partnership Project (3GPP)}, Technical Report (TR) 38.901, Dec
  2019, version 16.1.0.

\end{thebibliography}

\end{document}